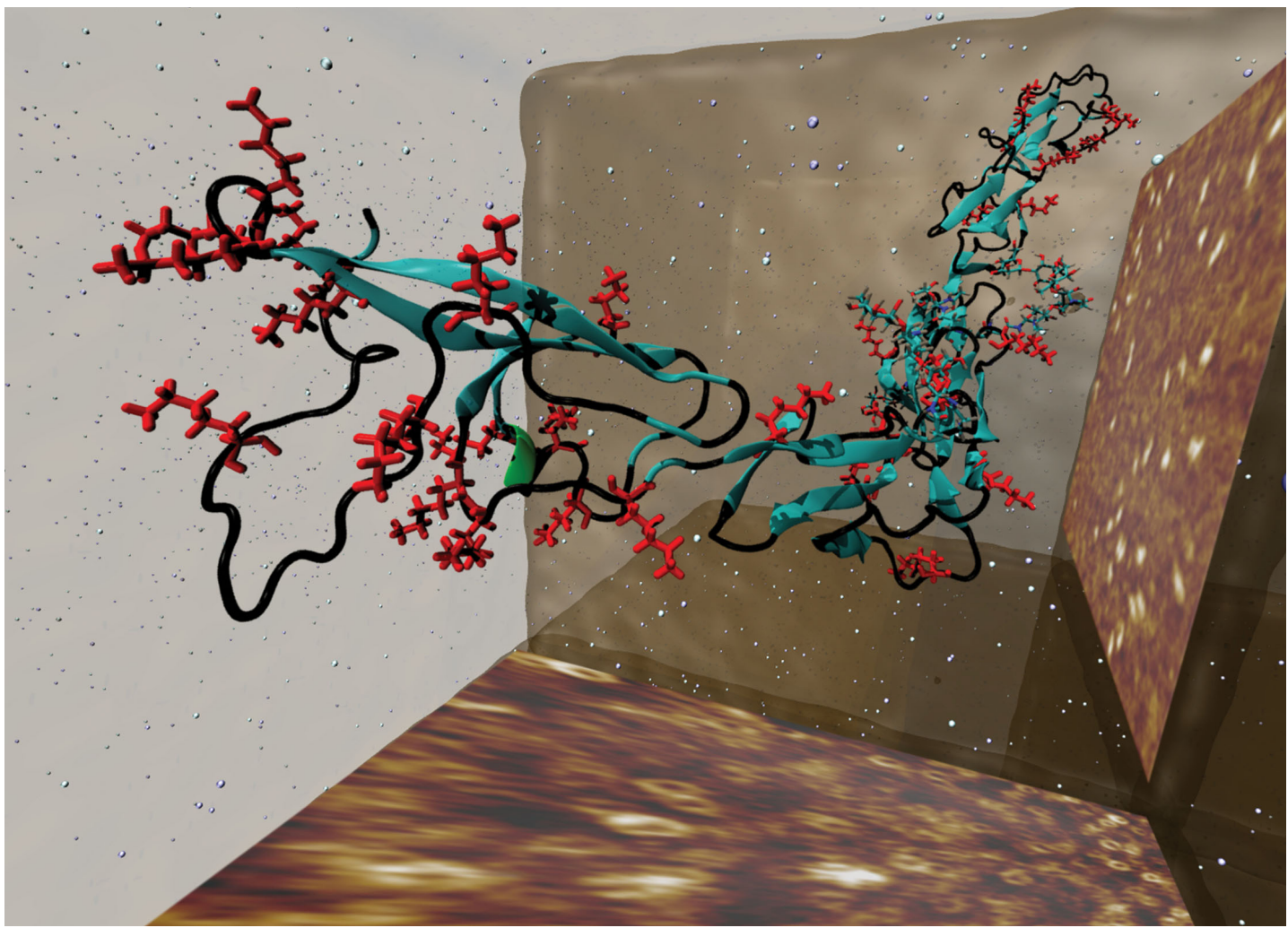

Showcasing research from the Group of Prof. Mihaela Delcea at the University of Greifswald, Germany

Lysine residues control the conformational dynamics of beta 2-glycoprotein I

Blood protein beta 2-glycoprotein I (beta2GPI) exhibits open and closed conformations and contains many lysine residues, marked in red. In the present work, the potential role of lysine in the conformational dynamics of beta2GPI is investigated. By chemical acetylation of lysine residues, the closed protein conformation opens up as revealed by atomic force microscopy images. Lysine plays a major role in stabilizing the beta2GPI closed conformation as confirmed by lysine charge distribution calculations.

As featured in:

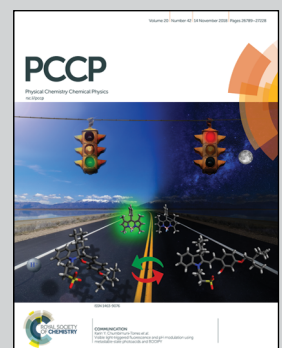

See Mihaela Delcea *et al.*, *Phys. Chem. Chem. Phys.*, 2018, **20**, 26819.

ROYAL SOCIETY OF CHEMISTRY

rsc.li/pccp

Registered charity number: 207890

# Lysine residues control the conformational dynamics of beta 2-glycoprotein I†

Ina Buchholz,[ab] Peter Nestler,[ab] Susan Köppen[cde] and Mihaela Delcea [ID] *[ab]

One of the major problems in the study of the dynamics of proteins is the visualization of changing conformations that are important for processes ranging from enzyme catalysis to signaling. A protein exhibiting conformational dynamics is the soluble blood protein beta 2-glycoprotein I (beta2GPI), which exists in two conformations: the closed (circular) form and the open (linear) form. It is hypothesized that an increased proportion of the open conformation leads to the autoimmune disease antiphospholipid syndrome (APS). A characteristic feature of beta2GPI is the high content of lysine residues. However, the potential role of lysine in the conformational dynamics of beta2GPI has been poorly investigated. Here, we report on a strategy to permanently open up the closed protein conformation by chemical acetylation of lysine residues using acetic acid *N*-hydroxysuccinimide ester (NHS-Ac). Specific and complete acetylation was demonstrated by the quantification of primary amino groups with fluoraldehyde *o*-phthalaldehyde (OPA) reagent, as well as western blot analysis with an anti-acetylated lysine antibody. Our results demonstrate that acetylated beta2GPI preserves its secondary and tertiary structures, as shown by circular dichroism spectroscopy. We found that after lysine acetylation, the majority of proteins are in the open conformation as revealed by atomic force microscopy high-resolution images. Using this strategy, we proved that the electrostatic interaction of lysine residues plays a major role in stabilizing the beta2GPI closed conformation, as confirmed by lysine charge distribution calculations. We foresee that our approach will be applied to other lysine-rich proteins (*e.g.* histones) undergoing conformational transitions. For instance, conformational dynamics can be triggered by environmental conditions (*e.g.* pH, ion concentration, post-translational modifications, and binding of ligands). Therefore, our study may be relevant for investigating the equilibrium of protein conformations causing diseases.

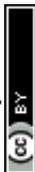

## Introduction

Beta 2-glycoprotein I (beta2GPI), previously named apolipoprotein H, circulates in blood at concentrations of 50–500 μg mL$^{-1}$,[1] hence it is one of the most abundant proteins in human serum. Although many details of the physiological function of beta2GPI are still unclear, it is known to interfere with agglutination mechanisms showing both anti-coagulation properties within the intrinsic coagulation pathway[2–4] and coagulant properties as negative feedback in extrinsic fibrinolysis.[2,5,6] Furthermore, beta2GPI comprises the major antigen for autoantibodies involved in antiphospholipid syndrome (APS),[7–9] which is one of the most common autoimmune diseases.

Beta2GPI is a single chain polypeptide with a molecular weight of about 50 kDa, including 326 amino acids.[10] It is organized in five domains (Fig. 1) that are connected by flexible random coil regions. Domains I (DI) to IV (DIV) consist of 60 residues each, showing repeating primary amino acid sequences and similar domain foldings.[11,12] Each of these domains includes beta-sheet (cyan) and random coil (black) secondary structures as well as two self-connecting disulfide bonds. Furthermore, DIII and DIV are highly glycosylated. In contrast, DV has two additional amino acid sequences: one section of six amino acids (Lys282–Lys287) within a short beta-hairpin, which is important for generating a lysine-rich positively charged patch. This patch is critical for binding negatively charged macromolecular structures, among other anionic phospholipids in membranes.[13–16] Another section of 20 amino acids comprises a C-terminal flexible loop (Phe307–Cys326) fixed by an additional disulfide bond, which also includes two short alpha-helical regions (green). From the total of 326 amino acids, more than 9% (30 residues, red colored) are lysine residues, which are mainly surface exposed. Half of them are distributed within the loop regions in DV,[11] giving this domain a highly positive charge.

[a] *Institute of Biochemistry, Ernst-Moritz-Arndt University Greifswald, 17489 Greifswald, Germany. E-mail: delcea@uni-greifswald.de*
[b] *ZIK HIKE – Zentrum für Innovationskompetenz, Humorale Immunreaktionen bei kardiovaskulären Erkrankungen, Ernst-Moritz-Arndt University Greifswald, 17489 Greifswald, Germany*
[c] *Hybrid Materials Interfaces Group, Faculty of Production Engineering, University of Bremen, 28359 Bremen, Germany*
[d] *Bremen Center for Computational Materials Science, 28359 Bremen, Germany*
[e] *MAPEX Center for Materials and Processes, University of Bremen, 28359 Bremen, Germany*

† Electronic supplementary information (ESI) available: S1: SDS-PAGE and anti-acetylated lysine antibody western blot of beta2GPI after acetylation. S2: CD spectra of different beta2GPI species. S3: dynamic light scattering data of beta2GPI species. S4: quantitative shape analysis after AFM imaging of beta2GPI. See DOI: 10.1039/c8cp03234c

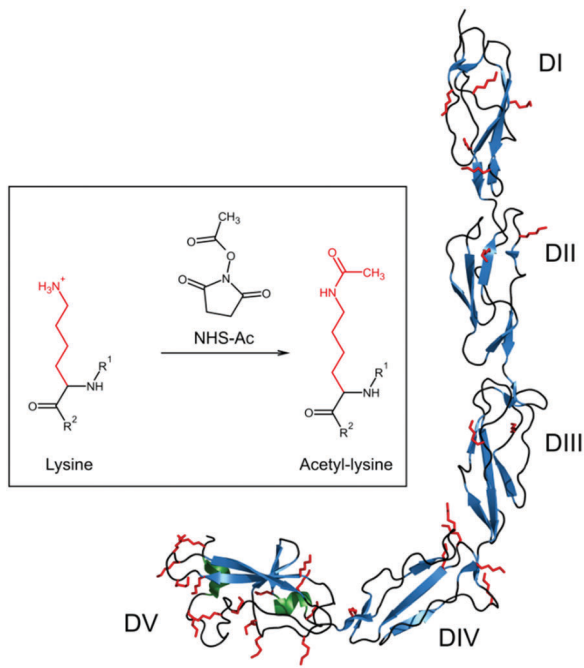

Fig. 1 Structure of beta2GPI in open conformation[11] (PDB-ID 1C1Z). The different secondary structure elements alpha-helix (green), beta-sheet (cyan) and random coil and turn regions (black) are shown. All lysine residues are marked in red as a stick model. The inset shows the reaction of lysine residues, which are positively charged at physiological pH, to uncharged acetyl-lysine by the acetylation reagent acetic acid N-hydroxysuccinimide ester (NHS-Ac). $R^1$ and $R^2$ refer to the continuation of the beta2GPI peptide chain. PyMOL 2.0 software was used to create this figure.

Beta2GPI is known to adopt different protein conformations.[17,18] A rather circular (closed) conformation, likely to be the inactive form circulating in blood, was visualized by Agar et al. and is characterized by the interaction of DI and DV.[19] A linear and fishhook-like (open) conformation was revealed by X-ray crystal structure analysis.[11,12] It has been shown that the open conformation exposes a hidden antibody-binding site called a "cryptic epitope" within DI–DII (i.e. Gly40–Arg43, Asp8–Asp9 and DI–II linker region[20]) and allows APS antibody binding that strongly correlates with thrombosis.[21–27] In addition, an S-shaped conformation, which is most likely an intermediate form, was described.[28]

The main question is what induces or facilitates the opening of beta2GPI. Possible triggers could be stress conditions (e.g. change in local pH or ion concentration), ligand binding,[29] or a post-translational modification of the protein. Krilis and coworkers showed that beta2GPI comprises different cysteine redox states,[30,31] but other post-translational modifications in beta2GPI have been poorly investigated.[32] Lysine residue acetylation via acetyltransferases is one of the most common post-translational modifications in proteins.[33,34] Lysine acetylation has also frequently showed an impact on immune system regulation.[35]

It was previously shown that beta2GPI opens up by shifting pH and salt concentration and that Lys305 and Lys317 located in DV are accessible to trypsin cleavage only in the open conformation.[19] The presence of a high content of lysine residues in DV, which might be involved in the opening of the protein upon binding to phospholipids, suggests that lysine may control protein conformation via electrostatic interactions between DV and DI. In this study, we report on a strategy to permanently uncharge beta2GPI lysine residues by acetylation and investigate their role in changing from the closed to open conformation using spectroscopic and imaging techniques.

## Experimental

Unless otherwise stated, all chemicals were purchased from Sigma (Taufkirchen, Germany). Human plasma beta2GPI was bought from Enzyme Research Laboratories (MW: 48 kDa; South Bend, USA). It is further referred to as the "untreated" species in Tris-buffered saline (TBS; 50 mM Tris, 0.15 M NaCl, pH 7.4). Protein concentration was determined using the bicinchoninic acid assay kit (Sigma).

#### Beta2GPI pH shift treatment

Following the protocol of Agar et al.[19] with slight modifications, 3 μM beta2GPI was dialyzed against 50 mM Tris and 1.15 M NaCl at pH 11.5 for 60 h and at 4 °C using 8 kDa cut-off dialysis cassettes (GE Healthcare, Freiburg, Germany). Afterwards, the samples were transferred to TBS solution using centrifugal filter units (UFC201024, Merck, Darmstadt, Germany) at 4000g and were analyzed within 6 h.

#### Beta2GPI acetylation

Acetic acid N-hydroxysuccinimide ester (NHS-Ac; Carbosynth Ltd, Berkshire, UK) dissolved in dimethylformamide (DMF) was added to 0.6 μM beta2GPI in phosphate-buffered saline (PBS; L1820, Merck) at pH 7.5 to reach NHS-Ac/lysine (mol/mol) ratios of 0, 1, 10, 100, and 1000, respectively. The content of DMF (0.6% v/v) was kept equal for all samples. The reaction mixture was incubated for 1 h at 25 °C under mild shaking. Afterwards, the samples were washed four times with PBS using centrifugal filter units (UFC501024, Merck) at 10 000g and finally transferred to TBS.

#### SDS-PAGE, native PAGE and western blot analysis

Beta2GPI samples were treated under reductive conditions and loaded into 4–12% gradient bis-Tris SDS gels (Thermo Fisher, Darmstadt, Germany). For native PAGE, the samples with 50 mM imidazole and 50 mM NaCl at pH 7.0 were visualized using 11% Tris–glycine native PAGE gels with SimplyBlue Coomassie stain (Thermo Fisher). For western blot analysis, the bands were blotted to a nitrocellulose membrane (Amersham Protran 0.2 μm NC, GE Healthcare) by semi-dry transfer for 30 min at 15 V. The membrane was blocked with 5% milk powder solution in TBS containing 0.05% Tween20 (TBST) for 1 h at room temperature (RT). Further, the membrane was washed three times with TBST and incubated overnight at 4 °C with anti-acetylated lysine antibodies (Cell Signaling Technology, Leiden, Netherlands) in 5% bovine serum albumin/TBST. After three washing steps, anti-rabbit IgG-coupled peroxidase antibody (Jackson ImmunoResearch Laboratories Inc., West Grove, USA) in 5% milk powder solution/TBST was incubated for 1 h at RT. After three additional washing steps, the membrane was stained with a peroxidase substrate (West Pico Chemiluminescent Substrate, Thermo Fisher).

### Quantification of primary amino groups

The content of primary amino groups was determined by using the fluoraldehyde o-phthalaldehyde (OPA) reagent (Thermo Fisher) in PBS. 20 μL of 2 μM acetylated beta2GPI samples (not transferred to TBS) and 20 μL of 6-aminohexanoic acid (concentration range: 0–300 μM) were placed on a black microtiter plate (SPL Life Sciences, Gyeonggi-do, Korea). 200 μL of the OPA reagent was added and the fluorescence ($\lambda_{ex}$ 340 nm and $\lambda_{em}$ 455 nm) was recorded within 5 min. 6-Aminohexanoic acid was plotted as a standard of known concentration to calculate the content of primary amino groups of the acetylated beta2GPI samples.

### Circular dichroism (CD) spectroscopy

CD spectra were acquired using a Chirascan CD spectrometer (Applied Photophysics, Leatherhead, UK) equipped with a temperature control unit (Quantum Northwest, Liberty Lake, USA) at 25 °C. The measurements in the far-UV region (195–250 nm) were performed using a 0.5 mm path length cuvette (106-QS; Hellma Analytics, Müllheim, Germany) and a protein concentration of 10 μM. CD spectra of the near-UV region (250–330 nm) were recorded using a 2 μM beta2GPI sample in a 10 mm path length cuvette (104-QS; Hellma Analytics). The spectra were acquired with a bandwidth of 1.2 nm, a scanning speed of 15 nm min$^{-1}$ and five repetitions. During data analysis, all spectra were blank corrected and the wavelength-dependent mean residue delta epsilon (MRDE) was calculated to normalize data for concentration $c$, the number of amino acids AA and the path length of the cuvette $d$ according to:

$$\text{MRDE} = \frac{\text{CD [mdeg]}}{c \text{ [M]} \cdot \text{AA} \cdot d \text{ [cm]} \cdot 32\,982}$$

### Dynamic light scattering (DLS)

Dynamic light scattering measurements were performed using a Zetasizer Nano-ZS (Malvern Instruments, Herrenberg, Germany). Beta2GPI samples (6–8.5 μM) were centrifuged at 10 000g for 10 min and the supernatant was vacuum-degassed for 30 min at 25 °C. Afterwards, 120 μL of each sample was transferred to a 10 mm path length cuvette (Brand, Wertheim, Germany), equilibrated to 25 °C for 5 min and measured 10 times at a detector angle of 173°. Each measurement consisted of 20 runs with a run duration of 10 s. A refractive index of 1.45 and an absorption of 0.001 with standard solvent parameters as referred to water were used. Data of the hydrodynamic diameter $D_H$ were analyzed using Zetasizer software 7.11 and are displayed as intensity vs. diameter size plot normalized to 100% intensity. The hydrodynamic diameter $D_H$ represents the diameter of a spherical shaped particle that diffuses with the same mobility as the beta2GPI molecule including the nearest shell of water molecules.[36]

### Atomic force microscopy (AFM)

AFM imaging was carried out using a Multimode atomic force microscope equipped with a Nanoscope IIIa controller (Veeco/Digital Instruments, Santa Barbara, USA). The AFM piezoelectric scanner was calibrated using the calibration gratings TGZ01 and TGG01 (MikroMasch, Tallinn, Estonia). Images were captured in AFM tapping mode using OMCL-AC160TS cantilevers with a radius of curvature of 10 nm and a spring constant of 42 N m$^{-1}$ (Olympus Corporation, Hamburg, Germany).

### AFM sample preparation

With a surface roughness of <0.1 nm, mica (Nanoscience Instruments, Phoenix, USA) served as the substrate for beta2GPI adsorption and AFM imaging. Prior to beta2GPI adsorption, mica sheets were freshly cleaved and treated according to the RCA standard procedure[37] in order to increase the negative surface charge density. The substrates were exposed to 1 μg mL$^{-1}$ (if not stated otherwise) beta2GPI samples in TBS with an adsorption time of 2 min to adjust for a surface coverage that allowed discrimination of individual proteins. Subsequently, the substrates were washed for one minute in deionized water (Milli-Q, Millipore, Billerica, USA) and dried in a laminar flow box (ScanLaf Class 2, LaboGene, Lynge, Denmark).

### AFM data processing and evaluation

AFM images were analyzed to determine the conformation of single, flatly adsorbed beta2GPI molecules. As a quantitative and unambiguous conformation criterion, we used the aspect ratio (length/width) of beta2GPI particles, which was determined by applying a home-written Matlab (The MathWorks, 2010b, Natick, USA) script. Image analysis was carried out in three steps: (1) the position of each protein on the substrate was localized by determining the local maxima. To account only for flatly adsorbed beta2GPI and to exclude contamination and substrate surface roughness, local maxima above 0.8 nm height and local maxima of less than 0.2 nm height were rejected. (2) Subsequently, the AFM height data were simplified into a binary map assigning every point to be associated either with the substrate or a particular beta2GPI molecule. For this purpose, the contour line surrounding a particular local maximum showing half the maximum height was defined as the lateral boundary of this particular beta2GPI molecule, analogous to the full width at half maximum (FWHM) in peak analysis. The area encircled by this boundary line corresponds to the area occupied by a certain beta2GPI molecule on the substrate. Particles with a lateral area above 200 nm$^2$ were excluded to account only for single, isolated beta2GPI molecules. (3) The molecular shape was further analyzed by numerically calculating the moment-of-area tensor for each localized beta2GPI. The moment-of-area tensor is defined as:

$$T = \begin{pmatrix} \int (y-y_0)^2 dy & \int (x-x_0)(y-y_0) dx dy \\ \int (x-x_0)(y-y_0) dx dy & \int (x-x_0)^2 dx \end{pmatrix}$$

where $x$ and $y$ are Cartesian coordinates in the substrate plane and where $x_0$ and $y_0$ are the coordinates of the protein's center of mass in the substrate plane.[38] The tensor $T$ is an integral quantity to measure the deformation of a two-dimensional object. The two eigenvalues ($\lambda_1, \lambda_2$) of $T$ are related to the length and width of the examined protein, respectively. Thus, the ratio $R = \max(\lambda_1, \lambda_2)/\min(\lambda_1, \lambda_2)$ is a measure for the aspect ratio length/width of a flatly adsorbed beta2GPI molecule. In the hypothetical case of a perfect circle, the aspect ratio equals $R = 1$. However, $R$ increases with an increasing degree of deviation from a circular shape. Beta2GPI molecules in a closed conformation are identified by an aspect ratio between $R = 1$ and $R = 3$, which was revealed by analyzing a set of beta2GPI molecules showing the characteristic torus shape. In contrast, beta2GPI molecules in an open conformation lead to a typical aspect ratio between $R = 3$ and $R = 10$. Hence, a threshold value of $R = 3$ was chosen to distinguish between beta2GPI in the open and closed conformations using automated script analysis. The $R$ value data of each analyzed beta2GPI particle were finally represented as box plots. Each box plot shows an independently prepared sample counting approximately 100 single beta2GPI molecules. The quantiles of the box refer to 25 and 75%, whereas the whisker includes the quantiles of 5 to 95% of the population. The percentage values of the closed and open protein conformations were determined from the total number of analyzed particles for each set of experimental conditions.

### Calculation of the distribution of titratable amino acids

The pH specific contribution of the most prominent titratable amino acids in beta2GPI was calculated with the H++ server version 3.2 (http://biophysics.cs.vt.edu/H++)[39] and the AMBER ff14SB force field. The internal dielectric was set to 10 and the external dielectric was set to 80. According to the experimental setup, salinities of 0.15 M at pH 7.4 and 1.15 M at pH 11.5 were specified. Within the total number of 326 amino acids in 1C1Z.pdb,[11] 93 titratable amino acids (glutamic acid, aspartic acid, lysine, histidine, arginine and tyrosine) and the N-terminus were taken into account.

### Calculation of the potential surface

The potential grid was calculated using the pdb2pqr server (http://nbcr-222.ucsd.edu/pdb2pqr_2.1.1/)[40] with the linearized Poisson–Boltzmann equation and 257 × 257 × 225 grid points. The ion concentrations of 0.15 M for pH 7.4 and 1.15 M for pH 11.5 were set according to the experimental setup.

## Results and discussion

To elucidate the role of lysine residues in shifting beta2GPI from the closed to open conformation, beta2GPI was specifically acetylated using acetic acid N-hydroxysuccinimide ester (NHS-Ac) as the acetyl donor[41,42] in order to permanently uncharge lysine residues. The reaction of lysine acetylation is given in the inset of Fig. 1. Beta2GPI was acetylated with ratios of 1, 10, 100, or 1000 NHS-Ac/lysine (mol/mol) to determine the optimum concentration of the acetylation agent needed for complete beta2GPI lysine amino acetylation.

### Native PAGE and western blot analysis of acetylated beta2GPI

Native PAGE analysis showed distinct bands of different beta2GPI acetylation species (Fig. 2A). Due to a decrease in positive charge density after reaction with increasing amounts of NHS-Ac, an increased electrophoretic mobility of the acetylated forms in comparison to the non-acetylated forms can be observed. The bands in lanes 1 and 6 refer to beta2GPI with the addition of the organic solvent dimethylformamide (DMF) and untreated beta2GPI, respectively. They display a comparable pattern and suggest that the DMF content does not alter the protein properties in native PAGE. A ratio of 1 NHS-Ac/lysine (mol/mol) (lane 2) already shows a slight difference in band mobility compared to lanes 1 and 6, indicating the initial acetylation of beta2GPI lysine residues. Gel bands become broader at ratios of 10 and 100 NHS-Ac/lysine (lanes 3 and 4, respectively), which suggests the presence of multiple acetylated beta2GPI species. With a ratio of 1000 NHS-Ac/lysine (lane 5), a more compact band appears, which accounts for the complete acetylation of beta2GPI existing as a homogeneous species. A reductive SDS-PAGE gel is shown in Fig. S1A (ESI†). The bands appear to be consistent and no protein aggregation or other complex formation can be observed upon acetylation reaction.

Western blot analysis of the native PAGE with a specific anti-acetylated lysine antibody was used to prove the presence and to estimate the amount of acetylated beta2GPI lysine residues.[43,44] As can be seen in Fig. 2B, no immunodetection signal is noticeable for untreated (lane 6), as well as DMF-treated (lane 1), and 1 NHS-Ac/lysine (mol/mol) treated beta2GPI (lane 2) residues. After treatment with a ratio of 10 NHS-Ac/lysine (lane 3), an initial acetylation is indicated by an immunodetection band matching the shape of the corresponding gel band in native PAGE (Fig. 2A, lane 3). Furthermore, NHS-Ac/lysine ratios of 100 and 1000 (lane 4 and 5, respectively) show a high binding of anti-acetylated lysine antibodies. These results demonstrate the presence of beta2GPI with different degrees of lysine residue acetylation. To compare the yield of acetylated beta2GPI, a western blot analysis with anti-acetylated lysine antibodies of the reduced SDS-PAGE was carried out (Fig. S1B, ESI†). It shows a similar behavior to the western blot of native PAGE. For non-acetylated or poorly acetylated beta2GPI, no immunodetection signal is found. After treatment with a ratio of 10 NHS-Ac/lysine, a faint band is visible. For ratios of 100 and 1000 NHS-Ac/lysine, strong signals with approximately the same intensity are detected, indicating minor differences in lysine acetylation between these acetylation ratios.

 

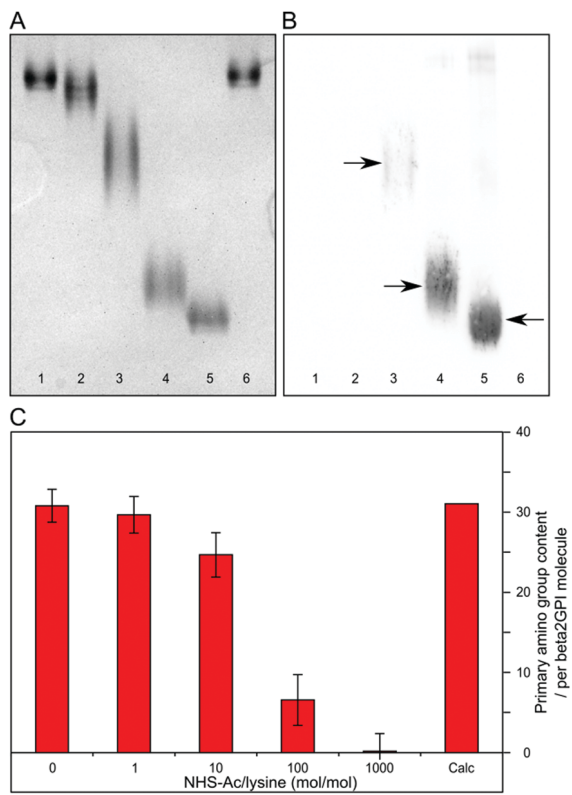

Fig. 2 (A) Native PAGE analysis of beta2GPI after acetylation with different molar ratios of NHS-Ac to lysine residues. Lane 1 refers to dimethylformamide (DMF)-treated beta2GPI. Lanes 2 to 5 refer to ratios of 1, 10, 100, and 1000 NHS-Ac/lysine (mol/mol), respectively. Lane 6 describes the untreated beta2GPI. (B) Western blot analysis of beta2GPI with anti-acetylated lysine antibodies after acetylation with different NHS-Ac/lysine ratios. Lanes are the same as described in (A). The bands corresponding to acetylated lysine residues of beta2GPI are highlighted by arrows. (C) Quantification of the primary amino group content after beta2GPI acetylation with different molar ratios of NHS-Ac/lysine (DMF-treated, 1, 10, 100, and 1000) determined by OPA assay. The means of four independent measurements as well as their standard deviations are shown. "Calc" refers to the calculated amount of primary amino groups (30 lysine residues plus the N-terminus) present in an untreated beta2GPI molecule.

### Determination of primary amino group content after beta2GPI acetylation

The content of acetylated lysine residues was quantified by using fluoraldehyde *o*-phthalaldehyde (OPA) assay, which provides a sensitive and specific reactivity towards primary amino groups.[45–47]

Fig. 2C shows the results of this assay for each NHS-Ac/lysine ratio used. The theoretically calculated value (Calc) of all primary amino groups within a beta2GPI molecule is 31 (30 lysine residues plus the N-terminus). The assay was tested for our system by comparing this calculated total amount of primary amino groups to the determined amount by OPA assay in untreated beta2GPI. In the absence of NHS-Ac, a primary amino group content of 31 ± 2 was observed. These values are comparable and render this assay suitable for our study. With a ratio of 1 and 10 NHS-Ac/lysine (mol/mol), a slight decrease in primary amino group content to 30 ± 2 (96%) and 25 ± 3 (79%) was observed, respectively. For an acetylation with a ratio of 100 NHS-Ac/lysine, a large decrease occurred (7 ± 3; 21%). Furthermore, after treatment with a ratio of 1000 NHS-Ac/lysine, almost no primary amino functions were present (0 ± 2). The lower amount of primary amino groups implies the increased formation of acetylated lysine residues. After the reaction with a ratio of 1000 NHS-Ac/lysine, beta2GPI was completely acetylated.

The results of the OPA assay and native PAGE together with immunodetection of acetylated lysine residues (western blot analysis) confirm that acetylation of beta2GPI is specific and complete at a molar ratio of 1000 NHS-Ac/lysine (mol/mol). A complete lysine residue acetylation was required to reach the potentially hidden lysine residues that could be involved in electrostatic stabilization of the closed beta2GPI conformation. Besides the acetylation of lysine residues, an acetylation of the N-terminal amino group could not be avoided. Although beta2GPI carries several glycosylation sites, none of them contains primary amino groups.

### Assessing conformational changes in beta2GPI

CD spectroscopy was used to assess whether beta2GPI preserves its structure after lysine acetylation. As a reference in our measurements, beta2GPI treated at pH 11.5 and a high salt content was included, which was proven by transmission electron microscopy (TEM) to be in the open conformation.[19] Far- and near-UV CD spectra of untreated (black), pH 11.5-treated (blue) and 1000 NHS-Ac/lysine (mol/mol) acetylated beta2GPI (red) are shown in Fig. 3A and B, respectively.

All beta2GPI species show the characteristic maximum at around 233 nm and a minimum at 201 nm, although acetylated beta2GPI displays a slight decrease of amplitude at a lower wavelength. Nevertheless, pH treatment and acetylation of lysine residues preserve the beta2GPI overall secondary structure. In addition, the CD profiles of different beta2GPI species do not change significantly within the near-UV region, suggesting an unmodified tertiary structure of acetylated beta2GPI.

The CD spectra of untreated and pH 11.5-treated beta2GPI showed the characteristic profile of beta2GPI, as published earlier.[48,49] After beta2GPI acetylation, a partial decrease in CD signal amplitude was observed. CD spectroscopy data of untreated compared to pH 11.5-treated beta2GPI suggest that secondary structural elements are preserved after the conformational change. Therefore, the change in the far-UV region of the acetylated beta2GPI probably results from the insertion of acetyl groups rather than a modified secondary structure by the opening of the protein. The integrity of the CD profile in the far-UV and near-UV region could be explained by the overall structural composition of the closed conformation compared to the open conformation of beta2GPI. Each of the five domains has a beta-sheet rich structure and is well separated from the next domain by a short but flexible random coil region (Fig. 1). These random coil regions may function as a hinge during the conformational change, and the secondary structural elements as well as tertiary structural arrangements remain unaffected. CD spectroscopy gives an overview of the structural properties of the beta2GPI species, but it is unable to identify specific structural features.

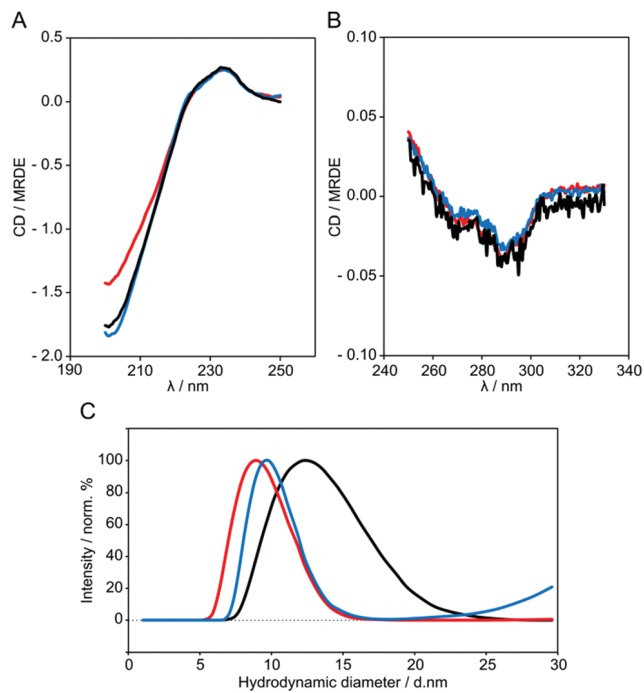

Fig. 3 (A) Far-UV and (B) near-UV regions of the CD spectra of untreated (black), pH 11.5-treated (blue) and lysine residue acetylated (red) beta2GPI with 1000 NHS-Ac/lysine (mol/mol) in TBS at pH 7.4 and at 25 °C. One representative CD spectrum for each beta2GPI species is shown. Far-UV and near-UV regions of the CD spectra were recorded with 10 μM and 2 μM beta2GPI in 0.5 mm and 10 mm path length cuvettes, respectively. (C) DLS data showing the hydrodynamic diameter $D_H$ (±SD of three independent measurements) of untreated beta2GPI (13.0 ± 0.9 nm, black), pH 11.5-treated (10.1 ± 0.2 nm, blue) and lysine residue acetylated beta2GPI with 1000 NHS-Ac/lysine (mol/mol) (9.5 ± 0.5 nm, red) measured in TBS at pH 7.4 and at 25 °C with 6–8.5 μM protein. One representative curve is shown for each beta2GPI species.

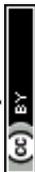

### Determination of protein size by dynamic light scattering (DLS)

DLS was used to determine the hydrodynamic diameter $D_H$ of untreated beta2GPI (black) as well as beta2GPI after pH 11.5 treatment (blue) and lysine residue acetylation (red) with a molar ratio of 1000 NHS-Ac/lysine (Fig. 3C). The hydrodynamic diameter assumes the particle to be of spherical shape and also includes the nearest shell of water molecules.[36] Untreated beta2GPI showed an increased $D_H$ ± SD of 13.0 ± 0.9 nm compared to the pH 11.5-treated (10.1 ± 0.2 nm) and lysine acetylated beta2GPI (9.5 ± 0.5 nm). Untreated beta2GPI is significantly larger than pH 11.5-treated and acetylated beta2GPI. This finding correlates with the native PAGE analysis (data not shown), where the gel bands of the pH 11.5-treated sample show a slightly increased electrophoretic mobility in comparison to untreated beta2GPI. Furthermore, the $D_H$ values are in agreement with Acquasaliente *et al.*[50] who found a $D_H$ of 8.7 ± 2.4 nm for open beta2GPI.

### Atomic force microscopy (AFM) imaging of beta2GPI

To further reveal a change in the beta2GPI conformation upon lysine acetylation, AFM imaging was used. As for the CD spectroscopy measurements, beta2GPI treated at pH 11.5 and a high salt content was used as a reference. Untreated beta2GPI showed the characteristic torus (donut-like) shape corresponding to the closed or circular conformation (Fig. 4A, left), as previously reported.[19] A local height minimum at the particle center was visible in close-up images of single beta2GPI molecules (Fig. 4A, middle). The lateral size dimension of these circular particles was determined to be 21 ± 6 nm in diameter. During statistical analysis with an automated script, the aspect ratio $R$ (particle length/width) was calculated for each beta2GPI particle of interest. Further, the median of $R$ for each sample consisting of approximately 100 beta2GPI particles is given (Fig. 4A, right). For untreated beta2GPI, the median of $R$ is 1.6 ± 0.1. A threshold of $R = 3$ was chosen to distinguish between the closed and open structures after analyzing a set of AFM close-up images, depicting isolated beta2GPI with characteristic torus and fishhook-shaped appearances, respectively. Untreated beta2GPI was quantified to be 93% in the closed conformation and 7% in the open conformation, indicating that the majority of particles are circular in shape. All values of the median of $R$ as well as the percentage fractions of open and closed proteins are summarized in Table 1.

After pH 11.5 treatment of beta2GPI, the AFM images show a mixture of circular structured as well as elongated, hook-shaped beta2GPI (Fig. 4B, left and middle). The elongated appearance is characteristic of beta2GPI in an open conformation, providing lateral size dimensions of 21 ± 4 nm × 10 ± 3 nm (length × width). These size dimensions have been deduced from the AFM cross-section data of beta2GPI in an open conformation by determining the end-to-end distance (length) as well as the thickness at the protein's narrowest point (width). The size dimensions provide information on a protein's maximum and minimum lateral extent. In contrast, $R$ is derived from the moment-of-area tensor $T$ of a particular beta2GPI. Since the tensor $T$ is an integral quantity of the area occupied by a certain beta2GPI, the resulting aspect ratio $R$ is robust to small fluctuations in the protein boundary contour line.[38] Irrespective of their conformation, flatly adsorbed beta2GPI molecules provided average height values of 0.5 ± 0.2 nm. The median of $R$ for the pH 11.5-treated samples was determined to be 3.8 ± 0.8 (Fig. 4B, right). Thus, after pH 11.5 treatment, the majority of beta2GPI could be identified as being in an open conformation. Quantitative analysis revealed that 62% of the population was in an open conformation and 38% was in a closed conformation. These values already show an increased proportion of open beta2GPI after reversible pH treatment.

The electrostatic interaction of lysine residues is disrupted by the loss of positive charges at pH 11.5, because the p$K_a$ of the lysine side chain amino group proton is around 10.5. The high salt content (1.15 M NaCl) during pH 11.5 treatment also shields the electrostatic interaction between positively charged lysine residues in DV and their interaction regions in DI–DII. However, this process is reversible with time and the beta2GPI conformation shifts back to a closed structure.

To confirm the influence of lysine residue acetylation on the beta2GPI conformation, the AFM imaging of beta2GPI after treatment with ratios of 10 and 100 (both given in Fig. S2, ESI†), and 1000 NHS-Ac/lysine (mol/mol) (Fig. 4C) was performed. The median of the aspect ratio $R$ (particle length/width) for 10 and 100 NHS-Ac/lysine was found to be 2.2 ± 0.2 and 2.8 ± 0.5, respectively. Single particle analyses revealed the percentage values of 75% closed and 25% open conformations for a ratio of 10 NHS-Ac/lysine, as well as 57% closed and 43% open beta2GPI structures for a ratio of 100 NHS-Ac/lysine. With a ratio of 1000 NHS-Ac/lysine, a complete lysine residue acetylation was achieved. Under these conditions (Fig. 4C, left and middle), AFM imaging also shows a mixture of both conformational states. The median of $R$ was determined to be 3.5 ± 0.3 (Fig. 4C, right).

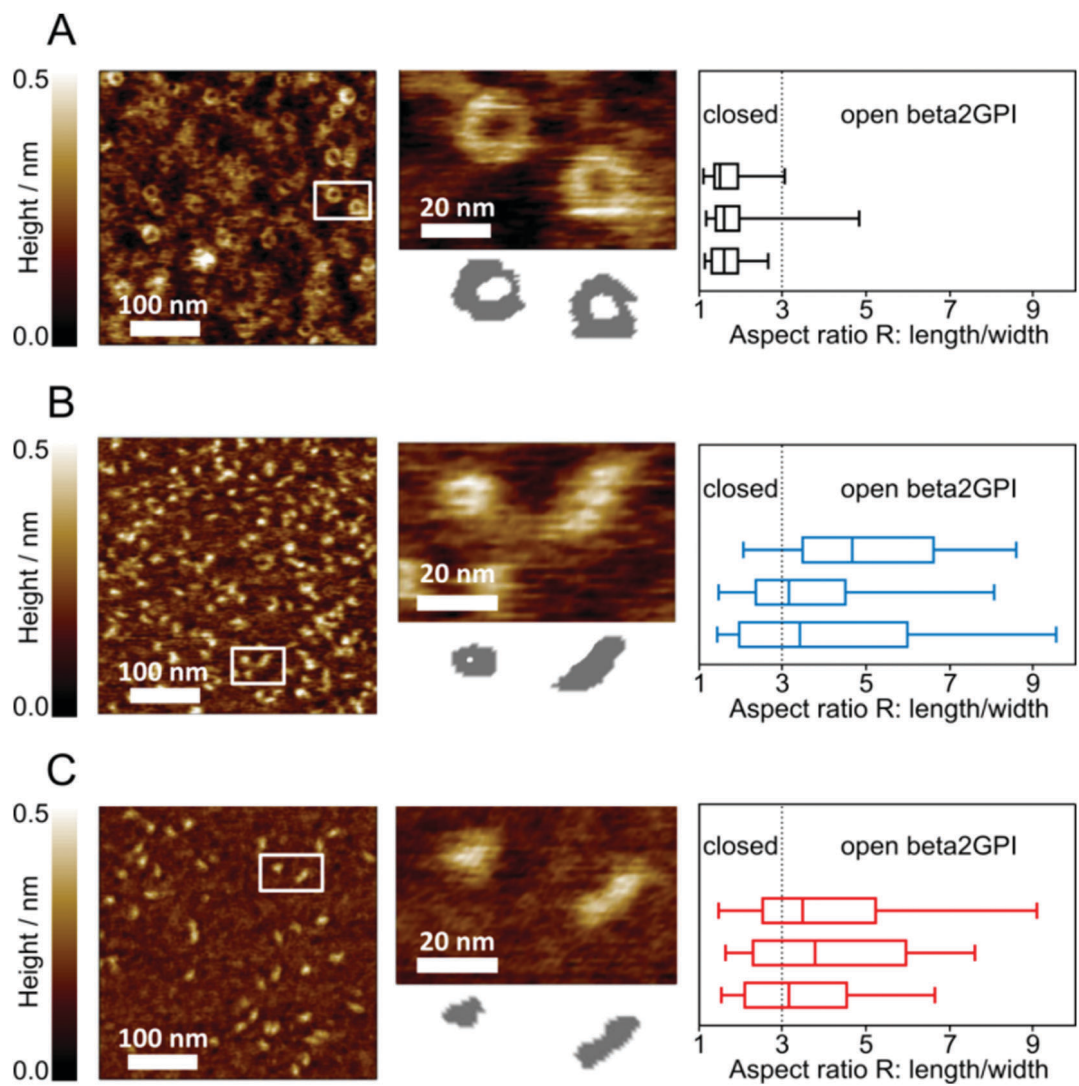

Fig. 4 AFM imaging of flatly adsorbed beta2GPI on a mica substrate. (A) Overview image of untreated beta2GPI adsorbed as 1 μg mL$^{-1}$ solution (left); close-up image of untreated beta2GPI in closed conformation (middle). (B) Overview image of beta2GPI after pH 11.5 treatment adsorbed as 1 μg mL$^{-1}$ solution (left); close-up image of two beta2GPI molecules in closed and open conformations, respectively (middle). (C) Overview image of beta2GPI after lysine residue acetylation with 1000 NHS-Ac/lysine (mol/mol) adsorbed as 5 μg mL$^{-1}$ solution (left); close-up image of two beta2GPI molecules in open and closed conformations, respectively (middle). For each close-up image, simplified representations of the molecular shapes (i.e. binary map diagram) are shown. The quantitative shape analysis of flatly adsorbed beta2GPI after AFM imaging is shown as the aspect ratio $R$ (particle length/width). Each box plot represents one independently prepared sample counting approximately 100 single beta2GPI molecules for untreated beta2GPI (A, right), beta2GPI after pH 11.5 treatment (B, right) and beta2GPI after lysine residue acetylation (C, right) with 1000 NHS-Ac/lysine (mol/mol). The threshold value to distinguish open from closed beta2GPI was $R$ = 3. Box plots show the median of $R$. Quantiles are set to 25 and 75%, whereas whiskers cover quantiles of 5 to 95% of the population.

Quantitative analysis showed an abundance (61%) of open beta2GPI compared to 39% closed proteins. The chemical acetylation of beta2GPI represents a permanent method to uncharge lysine residues and provides comparable results to the pH 11.5 treatment.

Although we have previously used AFM for imaging proteins in a liquid environment,[51,52] in this study, beta2GPI was dried on a mica substrate surface before imaging to avoid a decrease in image resolution. Furthermore, beta2GPI adsorbed flatly to the mica surface after drying, enabling conformation determination, whereas proteins in the upright orientation cannot be excluded in a liquid. However, drying beta2GPI on the mica substrate surface may influence the equilibrium of protein conformations. Another limitation could be the evaluation of particle shape by its appearance on the image. A minor percentage of particles may be miscalled as closed, because some open particles may adsorb as round shapes. Because of this, we rather consider the percentage of open particles as slightly underestimated. Moreover, we cannot distinguish between S-shaped and open or closed forms. Thus, for the simplicity of data evaluation, only open and closed conformations were accounted for.

Table 1 Summary of the median of $R$ and percentage fractions of closed/open beta2GPI. Quantitative shape analysis after AFM imaging of different beta2GPI species revealed the median of aspect ratio $R$ (particle length/width) with standard deviation (SD) of three independent sample preparations, as well as percentage fractions of closed and open proteins of untreated, pH 11.5-treated, and acetylated beta2GPI, respectively

| Beta2GPI species | Median of $R$ ($\pm$ SD) | Closed (%) | Open (%) |
| --- | --- | --- | --- |
| Untreated | 1.6 $\pm$ 0.1 | 93 | 7 |
| pH 11.5-treated | 3.8 $\pm$ 0.8 | 38 | 62 |
| 1000 NHS-Ac/lysine | 3.5 $\pm$ 0.3 | 39 | 61 |
| 100 NHS-Ac/lysine | 2.8 $\pm$ 0.5 | 57 | 43 |
| 10 NHS-Ac/lysine | 2.2 $\pm$ 0.2 | 75 | 25 |

The home-written automated script used to determine protein conformation allows objective and high-throughput analysis of hundreds of single proteins within a short time frame. Each single particle of interest is selected for calculating the aspect ratio $R$ (particle length/width). We chose a threshold $R = 3$ to distinguish between closed and open structures. This value was set after analyzing AFM close-up images depicting isolated beta2GPI molecules with characteristic torus or fishhook-shaped appearances. In comparison to Agar et al.,[19] we determined a lower percentage of beta2GPI in the open conformation after pH 11.5 treatment. Possible reasons for this may lie in our sample preparation procedure, which differs from the one reported before. In our opinion, the source and purification strategy of beta2GPI may play an important role in determining its susceptibility to undergoing conformational changes. Although the untreated beta2GPI shows more than a 90% closed conformation in both cases, the ability to change its structure during treatment could be different.

AFM imaging revealed particle lateral sizes for closed and open beta2GPI of 21 nm and 21 $\times$ 10 nm (length $\times$ width), respectively. However, imaging single molecules with AFM is always subjected to cantilever tip convolution,[53] because the molecular size and the cantilever tip curvature are within the same order of magnitude. As a result, the imaged beta2GPI molecules appear broadened and structural details of the proteins are potentially smoothed out. Therefore, these findings may result in an overestimation of beta2GPI particle size. However, because both DLS and AFM show protein sizes with a comparable order of magnitude, we can exclude the formation of protein aggregates on the mica substrate surface and assume that we address single adsorbed beta2GPI molecules in AFM imaging after drying. In addition, based on CD data, aggregation or denaturation of beta2GPI was excluded.

### Contribution of lysine residues to beta2GPI charge distribution

In order to theoretically prove that lysine acetylated beta2GPI mimics the amino acid charge distribution after pH 11.5 treatment, calculations of the distribution of the most prominent titratable amino acids at pH 7.4 compared to pH 11.5 were performed. According to the titration curves calculated with H++,[39] the total charge of beta2GPI varies from +6e at pH 7.4 to +21e at pH 11.5. The different total charges are the result of the proportion of titratable amino acids, as listed in Table 2. Lysine is present in all five domains, with a cluster of 16 residues located in DV. At pH 7.4 (Fig. 5A), all lysine residues are positively charged, whereas at pH 11.5 (Fig. 5B), 27 from the total 30 are neutralized. Three lysine residues, Lys202 and Lys210 (DIV) as well as Lys284 (DV), retain their positive charge due to micro-environmental conditions. Glutamic and aspartic acids are well distributed over all five domains (distribution not shown). Under physiological (pH 7.4) and basic pH (pH 11.5) conditions, all glutamic and aspartic residues are negatively charged. Arginine, tyrosine and histidine also do not change their protonation state within the pH range.

These findings prove that lysine residues change their protonation state significantly between pH 7.4 and 11.5 and could be the cause of the beta2GPI conformational change. Furthermore, fully lysine acetylated beta2GPI (1000 NHS-Ac/lysine) reflects pH 11.5 treatment in a good and irreversible way. In addition, the p$K_a$ of glycoside functional groups such as D-glucosamine, D-galactosamine and D-mannosamine lies between 7.8 and 8.5,[54] and therefore, they could change their protonation states within the given pH range. Nevertheless, all glycosides are aminoacetylated in the given structure.

In Fig. 5C and D, the isosurface at potential values of +1$k_B T$ (blue) and $-1k_B T$ (red) are shown. Due to the change in the protonation state of lysine residues between pH 7.4 and 11.5, the loss of a positively charged patch in DV can be observed. Apart from this fact, the proportions of isosurface at each pH do not differ remarkably. Since the development of the potential is known to be strongly salt concentration dependent,[56] we assume that the salt concentration of 0.15 M almost completely shields the electrostatic contributions in the ionic aqueous solution.

Based on these results, we conclude that the conformational change is more pH-dependent than salt dependent. To understand the beta2GPI conformational change in more detail and to estimate the interfaces of DI–II and DV interactions, further investigations focusing on molecular dynamic simulation studies are ongoing.

Table 2 Composition and proportion of protonation states of the most prominent titratable amino acids in beta2GPI (PDB-ID 1C1Z) calculated with H++[39]

| Amino acid[a] | Total no. | pH 7.4 | pH 11.5 |
| --- | --- | --- | --- |
| GLH/GLU | 20 | 0/20 | 0/20 |
| ASH/ASP | 14 | 0/14 | 0/14 |
| HIP/HIS | 5 | 0/5 | 0/5 |
| LYS/LYN | 30 | 30/0 | 3/27 |
| Prot./TYR | 14 | 0/14 | 0/14 |
| ARG/neutr. | 10 | 10/0 | 10/0 |

[a] The three-letter code illustrates the positively/neutral/negatively charged state of each amino acid residue, respectively. *E.g.* GLH is the protonated state of GLU.

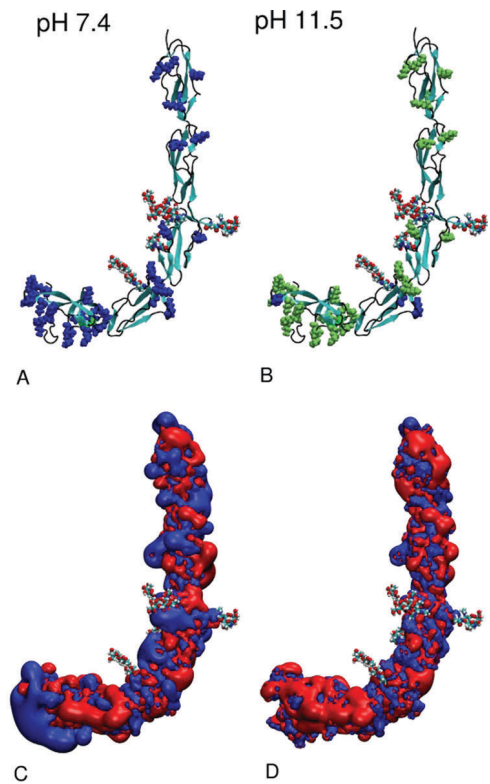

**Fig. 5** Schematic view of the structure of beta2GPI as provided by 1C1Z.pdb[11] under two different pH conditions: (A) 7.4 and (B) 11.5. The backbone is shown as a new cartoon model with a secondary structure code: beta-sheet (cyan), alpha-helix (green), random coil and turn regions (black). Glycosides linked to DIII and DIV are shown as vdW balls. Positively charged lysine residues are colored as blue and neutral lysine residues as lime vdW balls. In (C and D), the corresponding potential surfaces analogous to the experimental conditions are plotted with values of $+1k_{B}T$ (blue) and $-1k_{B}T$ (red). The graphics of this figure were created using VMD software.[55]

## Conclusions

In this study, we describe a simple strategy to permanently open up the closed conformation of the soluble blood protein beta2GPI by the chemical acetylation of lysine residues using acetic acid *N*-hydroxysuccinimide ester (NHS-Ac). Beta2GPI protein dynamics (shifting from closed to open conformation) was visualized for the first time by atomic force microscopy. We found that after lysine acetylation, the majority of proteins is in the open conformation. Using our approach, we confirmed that the electrostatic interaction of lysine residues plays a major role in stabilizing the beta2GPI closed conformation. Although a complete acetylation of lysine residues with a ratio of 1000 NHS-Ac/lysine (mol/mol) was achieved, not 100% of beta2GPI molecules are in open conformation. Lysine acetylation rather seems to shift the equilibrium between closed and open proteins. This might be explained by the fact that in addition to electrostatic interactions including positively charged lysine residues, other non-covalent interaction forces or external factors contribute to the stabilization of the closed conformation, *e.g.* involving a carbohydrate chain, as proposed by others.[17,28] Therefore, the closed form might be mainly, but not exclusively, mediated by the electrostatic interaction of beta2GPI lysine residues. Given that the beta2GPI open conformation is involved in the autoimmune disease antiphospholipid syndrome, the lysine residue acetylation of beta2GPI may have a significant biological impact causing the protein to become immunogenic (*i.e.* ability to induce an immune response) by adopting the open conformation to which disease-characteristic antibodies bind. Therefore, we foresee our simple *in vitro* approach for the chemical acetylation of lysine residues to be applied to understand disease mechanisms in which lysine-rich proteins (*e.g.* histones) undergoing conformational transitions are involved. Moreover, the conformational dynamics triggered by environmental conditions (*e.g.* pH, ion concentration, post-translational modifications, and binding of ligands), which may affect the equilibrium of protein conformations, can be visualized by atomic force microscopy (AFM). The combination of chemical lysine acetylation and AFM visualization could be applied in future studies to reveal mechanisms of diseases in which protein dynamics plays a major role.

## Conflicts of interest

There are no conflicts to declare.

## Acknowledgements

We thank Joost C. M. Meijers for fruitful discussion and suggestions regarding the pH treatment conditions of beta2GPI, Klaus Weisz for helping with CD data interpretation, and Walter Langel for discussion on molecular dynamics. We further acknowledge the support of Christiane A. Helm for providing access to AFM. We also thank Doreen Biedenweg for technical support with PAGE. Anisur Rahman and Thomas McDonnell are acknowledged for critical reading of the manuscript. This work was supported by the European Research Council (ERC) Starting Grant 'PredicTOOL' (637877).

## Notes and references

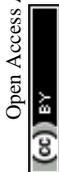